\documentstyle[preprint,aps,epsf]{revtex}
\tighten
\begin{document}
\draft
\title{Derivative expansions of the non-equilibrium effective action}
\author{Ian G. Moss}
\address{
Department of Physics, University of Newcastle Upon Tyne, NE1 7RU U.K.
}
\date{July 2001}
\maketitle
\begin{abstract}
New techniques for evaluating the closed time path effecive action for
non-equilibrium quantum fields are presented. A derivative expansion
is performed using a proper time kernel. Applications relevant to the
scalar field theory of thermal inflation are discussed and dissipation
terms resummed. The effective action of the electromagnetic field is
also considered. In this case the leading term can be related to the
conductivity of a plasma and has a simple interpretation in terms of
the classical Drude theory of conductivity.
\end{abstract}
\narrowtext

\section{introduction}

Many situations of interest in Particle Physics and Cosmology are
close to thermal equilibrium with fields which have slowly varying
ensemble averages. Two examples in the early universe are where the time
evolution is due to the expansion of the universe or the slow change
of an inflaton field. In is important in such situations to understand
both how dissipation affects the dynamics and how the how the dynamics,
in turn, reacts back on the thermal ensemble.

In quantum field theory, the behaviour of ensemble averages
can be usefully encoded in an effective action. For systems in
thermal equilibrium an imaginary time formulation is usually used
\cite{dolan,weinberg}, but this is only appropriate for slowly varying
spatial fields \cite{mosstoms,mosspol}. For systems that are not in
thermal equilibrium, the effective action can be introduced within the
closed time path formalism invented by Schwinger which allows ensemble
averages to evolve in time \cite{schwinger,keldysh}. The relativistic
formulation \cite{zhou,calzetta} allows a full account of the dynamics
of nonlinear quantum fields and thermal fluctuations.

The case of slowly varying ensemble averages lends itself to a
derivative expansion of the effective action. Terms that are linear
in time derivatives are responsible for dissipative effects, where
the short wavelength modes behave as a thermal bath damping the long
wavelength average. Early work was done on this problem by Hosoya et
al. \cite{hosoya}, who were studying the problem of reheating in the
inflationary universe. The first derivative terms in the closed time
path formalism have been studied for a self-interacting scalar field by
Gleiser et al \cite{gleiser}. Higher order derivative terms have been
obtained by Das and Hot \cite{das}, although they restricted themselves to
terms quadratic in the fields. Berera et al. \cite{berera} have further
examined the first derivative terms and done some preliminary work on
higher derivative terms. Their work has been largely stimulated by the
warm inflationary scenario, where the evolution of the inflaton field
is damped by thermal interactions \cite{moss1,berera1,berera2}.

The first problem in a derivative expansion is that free field theory is
not a good place to begin. This is  because a non-interacting system has
no way to re-establish thermal equilibrium after an external disturbance,
and so successive disturbances will drive the system further and further
from equilibrium. This problem could be avoided by considering only a
small time interval \cite{maria}, but the prefered method is to begin
with an equilibrium theory with a dressed propagator which damps down
the effects of the external forces \cite{gleiser,wang,parwani}. A new
justification of this approach based on the effective action will be
given in appendix A.

We shall see that the propagator has a simple representation in terms
of a proper time kernel. Unlike the zero temperature case, the kernel is
nonvanishing in both forwards and backwards proper time directions. The
use of the proper time kernel simplifies the derivative expansion of
the effective action.

Application of the derivative expansion to a scalar field coupled to
a heat bath confirms that the derivative expansion is an expansion in
powers of $\gamma^{-1}\partial_t$, where $\gamma$ characterises the
damping effects of the particles in the heat bath. Resummation of the
derivative expansion is possible in the case that the second and higher
derivatives of the averaged field vanish. The results indicate that the
resummation introduces logarithmic terms in the high temperature limit.

In an attempt to understand the necessity for including damping
terms in the derivative expansion the formalism is applied to Quantum
Electrodynamics in section V. Dissipative effects at non-zero temperatures
are well known physical phenomena. Constant electric fields, for example,
drive currents through the plasma. These can be calculated from the
nonequilibrium effective action. The current can also be estimated by
elementary Drude theory and demonstrates how the damping is related
to collisions between accellerated electrons and thermal electrons
or photons.

The quantum field theory conventions used are those of Weinberg
\cite{weinberg1}, in which $\hbar=c=1$, the metric signature is $-+++$
and the Feynman propagator is $(k^2+m^2-i\epsilon)^{-1}$.

\section{Closed time paths}

The closed time path formalism provides a definition of Green functions
and the effective action of a nonequilibrium quantum system. The path
integral version used here follows Calzetta and Hu \cite{calzetta}.

At the initial time $t_i$ the probability distribution of the quantum
field is given by a density matrix $\rho[\phi_1,\phi_2]$. The generating
functional is then defined by a path integral over the two fields $\phi_1$
and $\phi_2$ which are given the same values at the final time $t_f$,
\begin{equation}
Z[J_1,J_2]=\int d\mu[\phi_1]\int d\mu[\phi_2]\rho[\phi_1,\phi_2]
e^{iS[\phi_1]+i\int J_1\phi_1-iS^*[\phi_2]-i\int J_2\phi_2}
\end{equation}
where $S[\phi]$ is the classical action and two sources $J_1$ and $J_2$
have been introduced. The generating functional has an important symmetry,
\begin{equation}
Z[J_1,J_2]=Z[J_2,J_1]^*\label{symz}
\end{equation}
and, with suitable normalisation, $Z[J,J]=1$.

The generating functional is practically the same as any generating
functional for two scalar fields except for the special boundary
conditions. This similarity can be made more explicit by introducing a
metric $c_{ab}=\hbox{diag}(1,-1)$, then $J^2=-J_2$. The effective action
is therefore
\begin{equation}
\Gamma[\phi]= -i\log Z-J^a\phi_a
\end{equation}
where contracted indices $a$ indicate an integral over $x$ \cite{dewitt}
and $J^a$ is the solution to
\begin{equation}
\phi_a={\delta Z\over\delta J^a}.
\end{equation}
As usual,
\begin{equation}
{\delta\Gamma\over\delta\phi_a}=-J^a
\end{equation}
The ensemble average $\phi_c$ is defined by $\phi_c=\phi_1=\phi_2$,
which occurs when $J_1=J_2$.

The effective action can be approximated in various ways. A modified
version of the one loop approximation will be used here, where
$\Gamma\approx S+W$ and
\begin{equation}
W=\case1/2\log\det\left(1+\Sigma G_D\right).\label{cola}
\end{equation}
$\Sigma$ contains $\phi$ dependent mass insertions and $G_D$ is the
equilibrium ($J_a=0$) propagator. Specifically,
\begin{equation}
\Sigma^{ab}={\delta^2 S_I\over \delta\phi_a\delta\phi_b}
\end{equation}
where $S_I$ contains the interaction terms in the action and (in
momentum space)
\begin{equation}
(G_D^{-1})^a{}_b= (k^2+m_0^2)\delta^a{}_b+\Sigma_0{}^a{}_b\label{dressed}
\end{equation}
where $\Sigma_0$ is the sum of equilibrium self-energy corrections to
the propagator.

Using dressed propagators avoids divergent terms arising in the derivative
expansion \cite{gleiser,parwani}. In particular, it is the dissipation
introduced by the mass correction $\Sigma_0$ which balances the effect of
the external sources. A derivation of the corrected one loop approximation
(\ref{cola}) is given in appendix A.

Some general statements can be made about the form of the equilibrium
propagator $G$. It will be useful to review these properties of $G$
in order to introduce some matrix notation which will be used in subsequent
sections. 

The propagator can be obtained from the generating functional by 
\begin{equation}
G_{ab}(x,x')=-i\left.{\delta^2 Z\over\delta J^a(x)\delta J^b(x')}
\right|_{J_a=0}.
\end{equation}
The symmetry (\ref{symz}) of the generating functional implies (for
bosons)
\begin{eqnarray}
&&G_{11}(x,x')=G_{11}(x',x)=-G_{22}(x,x')^*=-G_{22}(x',x)^*\\
&&G_{12}(x,x')=G_{21}(x',x)=-G_{12}(x',x)^*.
\end{eqnarray}
Furthermore, there are two physical conditions. Causality implies
(see \cite{calzetta}),
\begin{equation}
G_{11}+G_{22}+G_{12}+G_{21}=0\label{causality}
\end{equation}
Thermal equilibrium imposes periodicity in the imaginary time direction,
\begin{equation}
G_{21}({\bf x},t,{\bf x'},t')=G_{12}({\bf x},t+i\beta,{\bf x'},t')
\end{equation}
Fermion propagators are antiperiodic.

The symmetry and the causality conditions reduce the number of independent
functions needed to describe the propagator. For example, it is possible to
write the propagator in terms of functions $G^>(x,x')$ and $G^<(x,x')$ 
\cite{gleiser}, 
\begin{eqnarray}
G_{11}(x,x')&=&+\theta(t-t')\,G^>(x,x')\,\,+\theta(t'-t)\,G^<(x,x')
\label{gbig}\\
G_{12}(x,x')&=&-\theta(t-t')\,G^>(x,x')^*+\theta(t'-t)\,G^<(x,x')\\
G_{21}(x,x')&=&+\theta(t-t')\,G^>(x,x')\,\,-\theta(t'-t)\,G^<(x,x')^*\\
G_{22}(x,x')&=&-\theta(t-t')\,G^>(x,x')^*-\theta(t'-t)\,G^<(x,x')^*
\end{eqnarray} 
where $G^>(x,x')=G^<(x',x)$. 

In thermal equilibrium, the propagator can also be expressed in terms of the
spectral function  
\begin{equation} 
\rho=iG_{21}-iG_{12}.
\end{equation}
In momentum space, the symmetries of the propagator imply that it
takes the form  
\begin{equation}
G^a{}_b=\case1/2\sigma\delta^a{}_b+\case{i}/2 N^a{}_b\,\epsilon\rho
\label{genp}
\end{equation}
where $\sigma=G_{11}-G_{22}$ and $\epsilon(\omega)$ is the sign function. 
The periodicity in imaginary time gives  
\begin{equation} 
N^a{}_b=\pmatrix{1\pm 2f&\mp 2f\cr2(1\pm
f)&-(1\pm2f)\cr}\epsilon, 
\qquad f(\omega)={1\over e^{\beta\omega}\mp
1}\label{nmatrix} 
\end{equation}
with the upper signs for bosons and the lower signs for fermions. Two
important properties of the matrix $N$ are $N^2=1$ and $N^a{}_b(-\omega)=
N_b{}^a(\omega)$.

The general form of the propagator can now be compared with the dressed
propagator (\ref{dressed}). We see immediately that the self energy must
have the matrix form
\begin{equation}
\Sigma_0{}^a{}_b=\delta m^2(k)\delta^a{}_b
-i\gamma(k)\, N^a{}_b
\end{equation}
where $\delta m^2$ and $\gamma$ are real functions. If we set
$m^2=m_0^2+\delta m^2$, then
\begin{equation}
(G_D^{-1})^a{}_b=(k^2+m^2)\delta^a{}_b-
i\gamma N^a{}_b\label{operator}
\end{equation}
Imposing $G_D^{-1}G_D=1$ enables us to deduce that
\begin{eqnarray}
\rho&=&i(k^2+m^2+i\epsilon\gamma)^{-1}-
i(k^2+m^2-i\epsilon\gamma)^{-1}\label{spectral}\\
\sigma&=&(k^2+m^2-i\gamma)^{-1}+
(k^2+m^2+i\gamma)^{-1}
\end{eqnarray}
The poles of the spectral function $\rho$ occur when $\omega=\pm\omega_k\pm
i\Gamma$, where $\omega_k=(|{\bf k}|^2+m^2)^{1/2}$ and
\begin{equation} 
\gamma(k)=2\omega\epsilon(\omega)\Gamma(k)\label{poles}
\end{equation}
The function $\Gamma(k)$ can be interpreted as a decay
width.

The propagator is often written in a hybrid form which depends on spatial
momentum and time \cite{gleiser}. The inverse transform can be evaluated by
using equations (\ref{genp}), (\ref{nmatrix}) and (\ref{spectral}), then for
$t>0$,
\begin{equation}
G^>({\bf k},t)=
{i\over 2\omega_k}\left(
f(\omega_k+i\Gamma)e^{i(\omega_k+i\Gamma)t}
+(1+f(\omega_k-i\Gamma))e^{-i(\omega_k-i\Gamma)t}\right)
\end{equation}
This is a very unwieldy expression to integrate, especially when there two or
more propagators involved in a perturbative diagram. An alternative method for
evaluating integrals of propagators is given in the next section.

\section{The proper time kernel}

A new idea will now be introduced to simplify manipulations
involving the propagator. This is the introduction of a proper time kernel,
related to the kernel which was introduced into zero temperature field theory
by Schwinger \cite{schwinger2}. It is also related to the use Feymann
parameters for the evaluation of Feynman diagrams.

The proper time kernel $K_{ab}(x,x',\tau)$ is a $2\times2$ matrix,
although the matrix indices will not be written out explicitly. For
convenience in switching between configuration and momentum space the bra and
ket notation will be used and the kernel regarded as an operator $K(\tau)$. It
satisfies a Schr\"odinger equation,
\begin{equation}
i\partial_\tau K=\Delta K+i\delta(\tau)\label{schrod}
\end{equation}
where $\Delta G=1$. The proper time kernel generates the propagator by
\begin{equation}
G(x,x')=i\int_{-\infty}^\infty d\tau\,\langle x'|K(\tau)|x\rangle\label{rep}
\end{equation}
Unlike at zero temperature, the proper time integral covers past and future
values.

For the dressed propagator $G_D$, the momentum space version of
$\Delta=\Delta_D$ is given by equation (\ref{operator}). There is a unique
kernel which satisfies (\ref{schrod}) and for which the integral (\ref{rep})
converges, 
\begin{equation}
K_D=\case1/2\left(N+\epsilon(\tau)\right)e^{-i\Delta_D\tau}
\end{equation}
In momentum space,
\begin{equation}
K_D(k,\tau)=A\,e^{-i(k^2+m^2)\tau}\label{kernel}
\end{equation}
where,
\begin{equation}
A=\case1/2\left(N+\epsilon(\tau)\right)e^{-\gamma|\tau|}
\label{amatrix}
\end{equation}
The matrix $N$ was given by equation (\ref{nmatrix}) in terms of the
distribution function $f$.

It is instructive to examine the behaviour of the kernel at the inverse
temperature $\beta\to\infty$. One finds
\begin{equation}
A\to\pmatrix{\theta(\tau)e^{-\gamma\tau}&0\cr
0&\theta(-\tau)e^{\gamma\tau}\cr}.
\end{equation}
where $\theta(\tau)$ is a step fuction. When substituted into (\ref{rep}) one
recovers the Feynman and Dyson propagators at zero temperature, with a
$i\gamma$ in the places where you would normally place the $i\epsilon$.

The non-eqilibrium proper time kernel for the operator
$\Delta=\Delta_D+\Sigma$ can be obtained by perturbation theory. The usual
expressions involving time ordered products cannot be used because of the
acausal nature of the propagator. Instead,  it is helpful to define the
convolution product for functions of $\tau$, 
\begin{equation} 
f\circ g=\int_{-\infty}^\infty d\tau' f(\tau-\tau')g(\tau'). 
\end{equation}
The kernel satisfies an integral equation 
\begin{equation}
K=K_D-iK_D\circ\Sigma K.
\end{equation}
The solution to the integral equation is
\begin{equation}
K=K_D\circ(1+i\Sigma K_D)^{-1}
\end{equation}
where repeated use of the convolution product is used in the expansion of the
inverse. The nonequilibrium propagator can be recovered from (\ref{rep}),
which allows us to write the corrected one loop effective action (\ref{cola})
as \begin{equation}
W={i\over 2}\int_{-\infty}^\infty d\tau\,
{\rm tr}\log\left(1+i\Sigma K_D\right).
\end{equation}
If the logarithm is expanded it is possible to use the composition
property $K_D\circ K_D=\tau K_D$ and the cyclic symmetries of convolutions to
write this in a form resembling Schwinger's zero temperature result,
\begin{equation}
W={i\over 2}\int_{-\infty}^\infty d\tau\,\tau^{-1}
{\rm tr}K+\hbox{constant}.
\end{equation}
The one loop divergences show up at small values of $\tau$, but the
proper time approach is especially well suited to trouble-free analytic
regularisation schemes \cite{dowker}.

To derive the derivative expansion of the effective action we suppose
that $\langle x'|\Sigma|x\rangle=\Sigma(x)\delta(x-x')$. The full expression
for a term $(\Sigma K_D)^n$ in a momentum space basis is 
\begin{equation} 
\langle k'|(\Sigma K_D)^n|k\rangle=
(-i)^n\,\Sigma(i\partial_k)K_D\circ\dots
\circ\Sigma(i\partial_k)K_D\,\delta(k-k')
\end{equation}
where $K_D\equiv K_D(k,\tau)$ on the right hand side. In configuration space,
\begin{equation}
\langle x|(\Sigma K_D)^n|x\rangle=
(-i)^n\int{d^4k\over (2\pi)^4}
\Sigma(x+i\partial_k)K_D\circ\dots\circ\Sigma(x+i\partial_k)K_D
\end{equation}
The derivative expansion is obtained from a Taylor expasion about
$\Sigma(x)$. However, we first introduce the expression (\ref{kernel})
for $K_D$. If we take $m^2$ to be independent of $k$ it is possible to
move the exponential factors around to leave,
\begin{equation}
\langle x|(\Sigma K_D)^n|x\rangle=
(-i)^n\int{d^4k\over (2\pi)^4}
e^{-im^2\tau-ik^2\tau/2}
\left(\Sigma(x+\delta)A\circ\dots\circ\Sigma(x+\delta)
A\right)e^{-ik^2\tau/2}\label{terms}
\end{equation}
where the $\delta$ in the $j$'th position along the convolution product
is the operator
\begin{equation}
\delta(\tau_j)=i\partial_k+2k\tau_j-k\tau\label{delta}
\end{equation}
It is convenient to adopt the notation \cite{moss}
\begin{equation}
\langle\dots\rangle=\int{d^4k\over (2\pi)^4}{\rm tr}\,
e^{-ik^2\tau/2}\dots e^{-ik^2\tau/2},\label{matnot}
\end{equation}
then the one loop action can be written in a compact form
\begin{equation}
W={i\over 2}\int d^4x\int_{-\infty}^\infty d\tau e^{-im^2\tau}
\langle \log\left(1+i\Sigma(x+\delta)A\right)\rangle
\end{equation}
where the terms expand to the full expressions (\ref{terms}).

A useful property of the one loop action is that the mass can be shifted
by an arbirary function $X(x)$,
\begin{equation}
W={i\over 2}\int d^4x\int_{-\infty}^\infty d\tau e^{-i(m^2+X)\tau}
\langle \log\left(1+i(\Sigma(x+\delta)-X)A\right)\rangle.\label{shift}
\end{equation}
This follows from the properties of the proper time kernel.

The effective field equations are obtained by differentiating the
effective action. The one loop contribution is therefore
\begin{equation}
{\delta W\over \delta \phi_a}=
-{1\over 2}\int_{-\infty}^\infty  d\tau e^{-im^2\tau}
\langle {\partial \Sigma\over\partial\phi_a}A
\left(1+i\Sigma(x+\delta)A\right)^{-1}\rangle\label{fieldeq}
\end{equation}
With the physical condition $\phi=\phi_1=\phi_2$, the function $\Sigma$
is related to the interaction Lagrangian ${\cal L}_I(\phi)$,
\begin{equation}
\Sigma^a{}_b={\delta^2 S_I\over\delta\phi_a\delta\phi^b}
=X\delta^a{}_b,\qquad
X={\partial^2{\cal L}_I\over \partial \phi^2}
\end{equation}
and we use the shifting property (\ref{amatrix}) to get the main result of this
section 
\begin{equation}
{\delta W\over \delta \phi_1}=
-{1\over 2}{\partial^3 {\cal L}_I\over\partial\phi^3}
\int_{-\infty}^\infty d\tau e^{-iM^2\tau}
\left\langle\case1/2\sigma_3A\left(1+i
\left(M^2(x+\delta)-M^2(x)\right)A\right)^{-1}\right\rangle\label{fe}
\end{equation}
where $M^2=m^2+X$ and $\sigma_3=\hbox{diag}(1,-1)$. 

The derivative expansion of the effective field equations is obtained by
expanding $M^2$ in powers of $\delta$ about $x$ and evaluating the resulting
matrix elements. The matrix $A$ was given in equation (\ref{amatrix}).

\section{Scalar Field Theory}

Now we turn to an explicit calculation based on a semiclassical
scalar field interacting with a heat bath of weakly interacting scalar
particles. The semiclassical field will be regarded as varying slowly in
time so as not to upset the thermal equilibrium of the heat bath. This
is an idealisation of the situation encountered in the critically damped
evolution of an inflaton field.

The choice of interaction Lagrangian for the semiclassical field $\phi$
and the heat bath fields $\eta$ will be
\begin{equation}
{\cal L}_I=\case1/2 g^2\Phi(\phi)\eta^2
\end{equation}
The self-interaction terms of both $\phi$ and $\eta$ will be assumed to
be of order $g^4$, which means that the one loop effective action comes
predominantly from the $\eta$ particles.

The dressed propagator in models of this type has been studied by Berera
et al. \cite{berera}. A reasonable approximation can be made by evaluating
the self energy at zero momentum. In the high temperature limit $T\gg m$,
for example,
\begin{eqnarray}
m^2&\approx&m_0^2+{g^2T^2\over 12},\\
\gamma&\approx&{g^4T^2\over 64 \pi}
\end{eqnarray}
for the theory with $\Phi=\phi^2$.

The one loop contribution to the effective field equations reads
\begin{equation}
{\delta W\over \delta \phi_1}=
-{1\over 2}g^2\Phi'
\int_{-\infty}^\infty d\tau e^{-iM^2\tau}
\left\langle\case1/2\sigma_3A
\left(1+ig^2\left(\Phi(x+\delta)-\Phi(x)\right)A\right)^{-1}\right\rangle.
\end{equation}
The shifting property of the effective action allows us to replace $m^2$
by $M^2(\phi)=m^2+g^2\Phi(\phi)$.

We shall try to recover all of the terms invoving
$\dot\Phi=\partial_t\Phi$ and neglect $\partial_t^2\Phi$ and higher
derivatives. Thus, since $\Phi_{,\mu}\delta^\mu=-\dot\Phi\delta_\omega$,
\begin{equation}
\left({\delta W\over \delta \phi}\right)_{1}=
-{1\over 2}g^2\Phi'\sum_{n=1}^\infty c_n
(g^2\dot\Phi)^n\label{series}
\end{equation}
where,
\begin{equation}
c_n=\int_{-\infty}^\infty d\tau e^{-iM^2\tau}i^n
\langle \case1/2\sigma_3 A(\delta_\omega A)^n\rangle\label{cn}
\end{equation}
The evaluation of the matrix element and proper time integral are
described in appendix B. To leading order in $\gamma$,
\begin{equation}
c_n=(-1)^n\int {d^3 k\over (2\pi)^3}{1\over
2\omega_k}{\partial_\omega^nf\over (2\gamma)^n}
\end{equation}
where $\omega_k^2=|{\bf k}|^2+M^2$. The term with coeficient $c_1$ agrees with
the results of Berera et al \cite{berera}.

The behaviour of the sum can be understood a little better when it is
expressed as an integral. To begin with, let $\dot\Phi$ be negative. Set
\begin{equation}
n(\mu)={1\over 4\pi^2}\int_0^\infty {dk\over e^{\beta(\omega-\mu)}-1}.
\end{equation}
The one loop term becomes
\begin{equation}
\left({\delta W\over \delta \phi}\right)_{1}=
g^2\Phi'\int_{-\infty}^0
n(\mu)e^{-2\gamma\mu/(g^2\dot\Phi)}d\mu\label{chain}
\end{equation}
Analytic continuation can be used for large positive $\Phi$, with a pole
at $\dot\Phi=2\gamma T/g^2$.

The integral can be further simplified in the high temperature limit
$T\gg M$. Using standard techniques for high temperature expansions gives
\begin{equation}
n(\mu)\sim -{1\over 8\pi^2\beta}\log\left(1-e^{\beta\mu}\right)
\end{equation}
for negative $\mu$. Inserting this into (\ref{chain}) gives
\begin{equation}
\left({\delta W\over \delta \phi}\right)_{1}=
-{g^2\Phi' T^2\over8\pi^2}{g^2\dot\Phi\over 2\gamma T}
\left(\psi\left(1-{2\gamma T\over g^2\dot\Phi}\right)-\psi(1)\right)
\label{resum}
\end{equation}
where $\psi$ is the digamma function. For small $\dot\Phi$, the result
gives
\begin{equation}
 \left({\delta W\over \delta \phi}\right)_{1}=
-{g^2\Phi' T^2\over16\pi^2}{g^2\dot\Phi\over 2\gamma T}
\log\left({g^2\dot\Phi\over 2\gamma T}\right)^2.
\end{equation}
For comparison, the first term in the series expansion (\ref{series})
gives a similar result but contains no logarithm of $\dot\Phi$. Beyond
the one loop approximation, the two-particle irreducible diagrams are
of order $g^4$, but the presence of a leading logarithm indicates that
resummation of the one loop approximation gives the largest contribution
to the effective action can be used consistently.

Higher derivative terms can be obtained by similar means. The simplest
term with two derivatives would be
\begin{equation}
\left({\delta W\over \delta \phi}\right)_2=
-{1\over 2}g^2\Phi'\int_{-\infty}^\infty d\tau e^{-im^2\tau}
\case{i}/2\langle\case1/2\sigma_3A\delta^\mu\delta^\nu A\rangle
X_{,\mu\nu}
\end{equation}
Using the methods described in appendix B, this reduces to
\begin{equation}
\left({\delta W\over \delta \phi}\right)_2=
-{g^4Z_1(T)\over 4\gamma^2}\Phi'\ddot\Phi
+{g^4Z_2(T)\over 24}\Phi'\Phi^\mu{}_\mu
\end{equation}
where $Z_1=-I_1[2\omega\partial f]$ and $Z_2=I_3[2f+1]$ in terms of the
integrals defined in appendix B. The second term agrees with earlier
results for spatial derivatives in reference \cite{mosstoms}, whilst for small
$\gamma$,
\begin{equation}
Z_1=4\beta\int {d^3k\over (2\pi)^3}f(1+f).
\end{equation}
with $f\equiv f(\omega_k)$. The first term is the dominant time-derivative
term for small $\gamma$. There are additional contributions from two-particle
irreducible diagrams at order $g^4$ according to the number of vertices, but
these diagrams give fewer inverse powers of $\gamma$ and do not affect the
leading term.

Berera et al. \cite{berera} have claimed that the leading term is
absent if the homogeneous limit (${\bf k}\to 0$) is taken after letting
$\Gamma\to 0$, as suggested by Gross at al. \cite{gross}. However, they
also find that the first derivative terms diverge when the calculation
is done in that order. In the method used here, all of the terms are
recovered even when the fields are not homogeneous. In the next section,
the situation in quantum electromagnatism is investigated to help
understand first derivative terms.

\section{Quantum Electromagnetism}

The effective action for the electromagnetic field at zero temperature
contains corrections from vacuum polarization. The proper time kernel
for a constant field can be obtained exactly \cite{schwinger2}, the
first correction to the effective action being the Euler Heisenberg term.

Dissipative effects prevent such a simple treatment of the nonzero
temperature problem. The leading term is due to the current induced
by the electric field, which will be calculated here from the derivative
expansion. The result will be compared to the electrical conductivity
calculated by elementary transport theory \cite{misner}.

The proper time kernel for electrons in a background elecromagnetic
field $F^{\mu\nu}$ satisfies a Schr\"odinger equation with the operator
\begin{equation}
\Delta=-(\partial+ieA)^2+m^2-\case1/2e\sigma_{\mu\nu}F^{\mu\nu}
\end{equation}
where $\sigma_{\mu\nu}=\case1/2i[\gamma_\mu,\gamma_\nu]$. Comparison with the
free theory shows that 
\begin{equation}
\Sigma=2eA_\mu k^\mu+e^2A_\mu A^\mu-\case1/2e\sigma_{\mu\nu}F^{\mu\nu}
\end{equation}

In thermal equilibrium, the electron propagator contains mass corrections
of order $e^2$, but the first disipative terms arise at order $e^4$. These
have been discussed in detail by Le Bellac \cite{lebellac}.

The corrected one loop contribution to the effective field equation is
obtained from (\ref{fieldeq}),
\begin{equation}
{\delta W\over\delta A^\mu}=
-{1\over 2}\int_{-\infty}^\infty d\tau e^{-im^2\tau}
\langle \case1/2\sigma_3\left(2ek_\mu+ie\sigma_{\mu\nu}k^{\nu}\right)
A\left(1+i\Sigma(x+\delta)A\right)^{-1}\rangle\label{em}
\end{equation}
In this case, since
\begin{equation}
{\delta \Gamma\over\delta A^\mu}=-J_\mu,
\end{equation}
the one loop term can be viewed as an induced electric current.

The calculations can be simplified by using Schwinger \cite{sg} gauge
in which
\begin{equation}
A_{\mu_1,\mu_2\dots\mu_n}={1\over n+1}F_{\mu_2\mu_1,\mu_3\dots\mu_n}
\end{equation}
To leading order,
\begin{equation}
\Sigma=eF_{\mu\nu}\delta^\mu k^\nu-\case1/2\sigma_{\mu\nu}F^{\mu\nu}
\end{equation}
The leading term can be evaluated using the formula given in appendix
B. The result is
\begin{equation}
{\delta W\over\delta A^\mu}=\sigma E_\mu
\end{equation}
where the electric field is $E=-F_{\mu}{}^0$ and, for small constant
damping $\gamma$, the conductivity
\begin{equation}
\sigma={e^2\over \gamma}\int {d^3 k\over (2\pi)^3}
{1\over e^{\beta\omega_k}+1}\label{sigma}
\end{equation}
with $\omega_k^2=|{\bf k}|^2+m^2$.

The result for $\sigma$ is of order $e^{-2}$. This can be understood
in terms of elementary Drude theory. In a plasma of density $n$, if
an electron accellerates for a time $\tau$ between collisions due to
the applied electric field and then moves off in a random direction,
the Drude conductivity will be
\begin{equation}
\sigma\approx ne^2\tau/m.
\end{equation}
The relaxation time $\tau$ can be infered from the location of the poles
of the dressed propagator $\omega_k+i\Gamma$. According to equation
(\ref{poles}), $\tau=\Gamma^{-1}=2m/\gamma$, which gives a result in
agreement with (\ref{sigma}). In terms of the collision cross section
$\sigma_T$, the relaxation time is roughly $1/(nv\sigma_T)$. Since
the cross section is of order $e^4$, the value of $\sigma$ is of order
$e^{-2}$.

\section{Conclusion}

We have seen how the derivative expansion of the nonequilibrium
effective action can be used to approximate the dynamics of a quantum
field interacting with thermal radiation. The importance of damping in
the thermal medium is very apparent and we have seen that the reason
for this can be understood in terms of scattering theory.

When the damping is small, the dominant time derivative terms at
order $n$ are always associated with $n$ inverse factors of the damping
coefficient $\gamma$. This implies that successive terms in the derivative
expansion will be decreasing in size if the $n$'th derivatives are of
order $\gamma^n$ or less. Exceptions can occur at phase transitions
where infra-red problems affect the momentum integrals. Furthermore,
resummations of the expansion series can lead to logarithmic terms,
as we saw for the power series in $\gamma^{-1}\dot\Phi$ in the high
temperature limit.

The derivative expansion for external electromagnetic fields in Quantum
Electrodynamics which was discussed in section V can easily be extended
to quadratic order where the results can be compared to zero temperature
results obtained from the Euler-Heisenberg Lagrangian. The derivative
expansion will give non-linear corrections to the conductivity in this
case. It should also include the Hall effect.

The proper time kernel was useful in helping simplify the derivative
expansions but it can also be applied in other contexts. It provides a
natural way to  extend the Feynman parameterisation of the propagator
to non-equilibrium Feynman diagrams which could then be used to
evaluate multi-loop diagrams. It would also be valuable
to have the exact proper time kernel in the presence of a constant
background electromagnetic field, since it is known at zero temperature
\cite{schwinger2}. Unfortunatly, this appears to be a highly non-trivial
problem.

\acknowledgments

I am indebted to Ian Laurie and Maria Asprouli for discussions on
non-equilibrium field theory.

\appendix

\section{The one loop effective action}

The one loop effective action can be derived from the generalised
effective action of Calzetta and Hu \cite{calzetta} (also developed in
refs. \cite{lawrie,amelino}). This form of the action uses a source
$J_{ab}$ quadratically coupled to the field $\phi_a$ to generate the
propagator $G_{ab}$. The effective action $\Gamma[\phi_a,G_{bc}]$ has
the perturbative expansion
\begin{equation}
\Gamma[\phi_a,G_{bc}]=S[\phi_a]-\case1/2\log\det(G_{ab})
+\case1/2\Delta^{ab}G_{ab}+\Gamma_2
\end{equation}
where $\Gamma_2$ is the sum of two-particle irreducible diagrams
(i.e. they cannot be disconnected by removing two lines) in a theory with
`shifted' vertices and
\begin{equation}
\Delta^{ab}[\phi]={\delta^2 S[\phi]\over\delta \phi_a\delta\phi_b}.
\end{equation}
The effective action $\Gamma[\phi_a]$ can be obtained by replacing the
propagator with $G_{bc}\equiv G_{bc}[\phi_a]$, obtained by solving
\begin{equation}
{\delta\Gamma\over\delta G_{ab}}=0
\end{equation}
which implies
\begin{equation}
(G^{-1})^{ab}=\Delta^{ab}+{\delta\Gamma_2\over\delta G_{ab}}.
\end{equation}
Hence
\begin{equation}
\Gamma[\phi_a]=S[\phi_a]-\case1/2\log\det(G_{ab})+E
\end{equation}
where $G_{ab}\equiv G_{ab}[\phi_a]$ and $E$ is a two-particle irreducible
remainder term.

The dressed equilibrium propagator $G_D$ is defined to be the value of
$G_{ab}$ when $\phi=0$, hence
\begin{equation}
(G^{-1})^{ab}=(G_D^{-1})^{ab}+\Sigma^{ab}+E^{\prime ab}
\end{equation}
where $\Sigma=\Delta[\phi]-\Delta[0]$ and $E'$ is another remainder
term consisting of one-particle irreducible diagrams which vanish when
$\phi=0$. The corrected one loop approximation consists of dropping $E$
and $E'$,
\begin{equation}
\Gamma[\phi_a]\approx S[\phi]+\log\det(1+\Sigma G_D)
\end{equation}
The error terms have at least two shifted vertices. In $\lambda\phi^4$
theory, they would be of order $\lambda^2\phi^2$ at most, whereas the
corrected one loop term is still valid for $\lambda\phi^2$ large. Some
examples of graphs contributing and not contributing to the one loop
approximation are shown in figure 1.

\begin{figure}
\begin{center}
\leavevmode
\epsfxsize=30pc
\epsffile{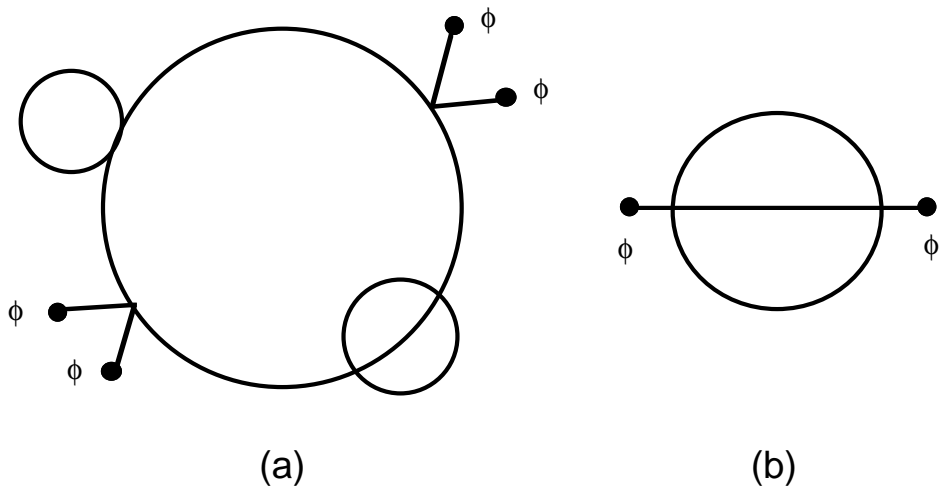}
\end{center}
\caption{Feynman diagrams for a scalar field with $\lambda\phi^4$
interactions. (a) shows an example of a diagram contributing to
the dressed one loop approximation used in the text; (b) shows a
two-particle irreducible diagram with shifted vertices not included at
one loop.}\protect\label{fig1}
\end{figure}

\section{Useful formulae}

The derivative expansion produces matrix elements containing
products of convolutions, derivatives and matrices. These can be evaluated
with the methods described below. To begin with, consider the
proper time convolutions. The convolution product is both associative and
cyclic. We can usefully define operators $d$ and $d^\dagger$ acting on
functions of proper time  
\begin{equation} 
df(\tau_i)=\tau_i f(\tau_i),\qquad f\circ d^\dagger g=(df)\circ g.
\end{equation}
The operator $d$ satisfies $d(f\circ g)=(df)\circ g+f\circ dg$.

Derivatives with respect to momentum appear in the operators $\delta$
(\ref{delta}). These operators can be expressed in terms of $d$ and
$d^\dagger$ by 
\begin{equation}
\delta={i\over\tau}\left(d^\dagger D-D^\dagger d\right)\label{delta2}
\end{equation}
where
\begin{equation}
D=\partial_k+ik\tau,\qquad D^\dagger=-\partial_k+ik\tau.
\end{equation}
Note also that $k=-\case1/2i(D+D^\dagger)/\tau$. The operators $D$ and
$D^\dagger$ behave like anihilation and creation operators inside the matrix
elements (\ref{matnot}), and satisfy 
\begin{equation}
[D^\mu,D^{\dagger\nu}]=2i\tau \eta^{\mu \nu}
\end{equation}
The strategy for evaluating the matrix elements involves moving the $D$
operators to the right and the $D^\dagger$ operators to the left.

The proper time kernel depended on a matrix $A$ defined in equation
(\ref{amatrix}). Convolutions of $A$ can all be obtained from the formula
\begin{equation}
d^mA\circ d^n A={n! \,m!\over (n+m+1)!}d^{n+m+1}A\label{comba}
\end{equation}
As a simplification we shall take the damping function $\gamma$ to be
constant. This can be justified as long as we are interested in only
the leading term in the limit $\gamma\to0$. (Derivatives of $\gamma$
can be retained if required.) Derivatives of $A$ can be then be dealt with
by using the formula,
\begin{equation}
A^m\circ(\partial_\omega^pA)\circ A^n=
\left(
{N^m(\partial_\omega^pN)N^{n+1}\over 2(2\gamma)^{n+m}}
S_{nm}(Nd)A-S_{mn}( Nd)A
{N^m(\partial_\omega^pN)N^{n+1}\over 2(2\gamma)^{n+m}}
\right)\label{longa}
\end{equation}
where
\begin{equation}
S_{nm}(x)=\sum_{r=0}^{n}\pmatrix{n\cr r\cr}
{(n+m-r)!\over n!\,m!}\,(2\gamma x)^r
\end{equation}
These formulae reduce the matrix products to expressions of the form
$\partial_\omega^m N\,d^nA$ and $d^nA\,\partial_\omega^m N$. The symmetry under
$\omega\to-\omega$ implies
\begin{equation}
\langle\sigma_3(\partial_\omega^p N)N^{p+1}S(Nd) A\rangle=
-\langle\sigma_3S(Nd)A(\partial_\omega^p N)N^{p+1}\rangle
\end{equation}
where $S(x)$ is any polynomial. After substituting for the matrices $A$
and $N$,
\begin{equation}
\langle\sigma_3(\partial_\omega^p N)N^{p+1}S(Nd) A\rangle=
2(-1)^{p}\langle(\partial^p_\omega f)\epsilon^{p+1}S(-\partial_\gamma)
\theta_+\rangle
\end{equation}
where $\theta_+(\tau)=e^{-\gamma|\tau|}$. 

Now consider the proper time integrals contained, for example, in equation
(\ref{fe}). These produce integrals of the form
\begin{equation}
\int_{-\infty}^\infty d\tau (i\tau)^{s-1}e^{-im^2\tau}
\langle\case1/2\sigma_3(\partial_\omega^n N) N^{n+1} S( Nd) A\rangle=
(-1)^nS(-\partial_\gamma)I_s[\partial_\omega^nf]\label{intm}
\end{equation}
where
\begin{equation}
I_s[h]=\int_{-\infty}^\infty d\tau (i\tau)^{s-1}\int {d^4k\over (2\pi)^4}
e^{-i(k^2+m^2)\tau}\theta_+(\tau)h(\omega).
\end{equation}

The integral $I_s$ can be evaluated by noticing that
\begin{equation}
\int_{-\infty}^\infty d\tau e^{-i(k^2+m^2)\tau}\theta_+(\tau)
=\epsilon(\omega)\rho(k)
\end{equation}
where $\rho$ is the spectral function (\ref{spectral}). For $s>2$, $I_s$
can be obtained by differentiation with respect to $m^2$. For $s\le2$,
analytic continuation must be used to obtain,
\begin{equation}
I_s[h]={-\pi^{3/2}\over \Gamma(\case5/2-s)}
{1\over 2\pi}\int {d^4k\over (2\pi)^4}|{\bf k}|^{2(1-s)}
\epsilon(\omega)\rho(k)h(\omega)
\end{equation}
This result is valid for arbitrary $\gamma$. A useful property of the
result is that $I_s[\partial_\omega h]=-I_{s+1}[2\omega h]$.

In the limit $\gamma\to0$, the spectral function reduces to a delta
function and
\begin{equation}
I_s[h]={\pi^{1/2}\over \Gamma(\case5/2-s)}
\int {d^3k\over (2\pi)^3}|{\bf k}|^{2(1-s)}{
h(\omega_k)\over 2\omega_k}
\end{equation}
with $\omega_k^2=|{\bf k}|^2+m^2$.

As an example, consider the evaluation of
\begin{equation}
M_n=\langle\case1/2\sigma_3 A\circ (\delta_\omega A)^n\rangle
\end{equation}
The $\delta$ can be expressed in terms of $D$ using equation
(\ref{delta2}). When moving the $D$'s to the right, they can contract
with other $D$'s or act on the $A$ matrices. For small $\gamma$, the
leading contribution comes when all of the $D$'s act on a single $A$
matrix. After using equation (\ref{comba}),
\begin{equation}
M_n=\sum_{j=0}^n (-1)^{n-j}\left({i\over 2\tau}\right)^n
{(2j)!(2n-2j)!\over j!(n-j)!}
\langle\case1/2\sigma_3 A^{2j}\circ (\partial^n_\omega A)\circ
A^{2(n-j)}\rangle
\end{equation}
Now equation (\ref{longa}) can be used. The following summation is needed,
\begin{equation}
\sum_{j=0}^n (-1)^j{(2j)!(2n-2j)!\over j!(n-j)!}
S_{2(n-j)\,2j}(x)=(4\gamma x)^nS_n(x)
\end{equation}
where
\begin{equation}
S_n(x)=\sum_{r=0}^n {1\over r!}(\gamma x)^r
\end{equation}
Hence,
\begin{equation}
M_n=(2i\gamma)^{-n}\langle \case1/2\sigma_3 (\partial^n_\omega N)N^{n+1}
S_n(Nd)A\rangle
\end{equation}
The proper time integral of $M_n$ used to define the coefficients $c_n$
(\ref{cn}) can be read off from equation (\ref{intm}), which reduces to
$c_n=(-1)^nI_1[\partial_\omega^nf]$ in the limit $\gamma\to0$.


\begin{references}
\bibitem{dolan}L Dolan and R Jackiw, Phys. Rev. D {\bf 9}, 3320 (1974)
\bibitem{weinberg}S Weinberg, Phys. Rev. D {\bf 9}, 3357 (1974)
\bibitem{mosstoms}I G Moss, D J Toms and A Wright, Phys. Rev. D {\bf 46}
1671 (1992)
\bibitem{mosspol}I G Moss and S J Poletti, Phys. Rev. D {\bf 47}, 5477
(1993)
\bibitem{schwinger}J Schwinger, J Math Phys {\bf 2}, 407 (1961)
\bibitem{keldysh}L V Keldysh, Zh. Eksp. Teor. Fiz. {\bf 47}, 1515 (1964)
\bibitem{zhou}G Z Zhou, Z B Su, B L Hao and L Yu, Phys. Rep. {\bf 118},
1 (1985)
\bibitem{calzetta}E Calzeta and B L Hu, Phys. Rev D {\bf 37}, 2878 (1988)
\bibitem{hosoya}A Hosoya and M Sakagami, Phys. Rev D {\bf 29}, 2228 (1984)
\bibitem{gleiser}M Gleiser and R O Ramos, Phys Rev. D {\bf 50}, 2441
(1994)
\bibitem{das}A Das and M Hott, Phys. Rev. D {\bf 50} 6655 (1994)
\bibitem{berera}A Berara, M Gleiser and R O Ramos, Phys Rev D {\bf 58},
123508 (1998)
\bibitem{moss1}I G Moss, Phys. Letts. B {\bf 154}, 120 (1985)
\bibitem{berera1}A Berera, Phys. Rev. Lett. {\bf 75}, 3218 (1995)
\bibitem{berera2}A Berera, Phys Rev D {\bf 54}, 2519 (1996)
\bibitem{maria}M Asprouli and V G Gonzalez, {\em Proceedings of the 5th
International Workshop on Thermal Field Theory} Regensburg, Germany,
(1998), hep-ph/9811469.
\bibitem{wang}E Wang, U Heinz and X Zang, Phys. Rev. D{\bf 53}, 5978
(1996)
\bibitem{parwani}P R Parwani, Phys. Rev. D{\bf 45}, 4695 (1992)
\bibitem{weinberg1}S Weinberg, {\em The Quantum Theory of Fields, Vol. 1}
(Cambridge University Press, Cambridge UK 1995)
\bibitem{dewitt}B. S. DeWitt, {\em Dynamical Theory of Groups and Fields}
(New York: Gordon and Breach 1965)
\bibitem{schwinger2}J S Schwinger, Phys. Rev. D {\bf 82} 664 (1951)
\bibitem{dowker}J S Dowker and R Critchley, Phys. Rev D {\bf 13} 3224
(1976)
\bibitem{lawrie}I D Lawrie, Phys. Rev. D {\bf 40}, 3330 (1989)
\bibitem{amelino}G Amelino-Camelia and S-Y Pi, Phys. Rev. D {\bf 47}
2356 (1993)
\bibitem{sg}J S Schwinger, {\em Sources, Particles and Fields}
(Addison-Wesley, New York 1973)
\bibitem{moss}I. G. Moss and W. Naylor, Class. Quantum Grav. {\bf 16}
(1999) 2611
\bibitem{gross}D J Gross, R D Pisarski and L G Yaffe, Rev. Mod. Phys. {\bf
53} 43 (1981)
\bibitem{misner}C W Misner and D H Sharp, Phys Letts {\bf 15} 279 (1965)
\bibitem{lebellac}M Le Bellac, {\em Thermal Field Theory} (Cambridge
University Press, Cambridge UK 1996)
\end{references}
\end{document}